\documentclass[10pt]{article}
\usepackage{indentfirst}
\usepackage[letterpaper, margin=1in]{geometry}
\usepackage{graphicx}
\usepackage{ragged2e}
\usepackage{hyperref}

\def\Title#1{\begin{center} {\Large #1 } \end{center}}
\def\Author#1{\begin{center}{ \sc #1} \end{center}}
\def\Address#1{\begin{center}{ \it #1} \end{center}}

\newcommand\pubblock{\rightline{\begin{tabular}{l} Proceedings of Lepton Photon 2025\\ \pubnumber\\
         \pubdate  \end{tabular}}}

\newenvironment{Abstract}
  {\begin{quotation}
   \begin{center}\large ABSTRACT\end{center}\bigskip}
  {\end{quotation}}

\newenvironment{Presented}{\begin{quotation} \begin{center} 
             PRESENTED AT\end{center}\bigskip 
      \begin{center}\begin{large}}{\end{large}\end{center} \end{quotation}}

\def\beq{\begin{equation}}
\def\eeq#1{\label{#1}\end{equation}}
\def\eeqn{\end{equation}}

\def\beqa{\begin{eqnarray}}
\def\eeqa#1{\label{#1}\end{eqnarray}}
\def\eeqan{\end{eqnarray}}

\let\bar=\overbar

\def\Dslash{\not{\hbox{\kern-4pt $D$}}}
\def\dslash{\not{\hbox{\kern-2pt $\del$}}}

\def\msb{{\bar{\ssstyle M \kern -1pt S}}}

\usepackage{color}
\usepackage{amsmath}

\newcommand\pubnumber{ CMS CR-2025/275 }

\newcommand\pubdate{November 26, 2025}

\def\affiliation{
On behalf of the CMS Collaboration, \\
Department of Physics,\\
Brown University, Providence, RI 02912, U.S.A}

\begin{document}

\large
\begin{titlepage}
\pubblock

\vfill
\Title{ Recent results on vector-like quarks and excited fermions at CMS }
\vfill

\Author{ Xiaohe Shen }
\Address{\affiliation}
\vfill
\begin{Abstract}
The most recent results on the searches for the vector-like quarks and excited fermions from the CMS Collaboration are presented. These results are based on the full Run 2 $\sqrt{s} = $13 TeV proton-proton collision data collected by the CMS Collaboration at the LHC from 2016 to 2018, which corresponds to an integrated luminosity of 138 fb$^{-1}$. No significant excess above the Standard Model expectation is observed. Exclusion limits are set at the 95\% confidence level on various benchmark models.
\end{Abstract}
\vfill

\begin{Presented}
32nd International Symposium \\
on Lepton Photon Interactions at High Energies \\
University of Wisconsin–Madison, Wisconsin, U.S.A \\ 
August 25-29, 2025
\end{Presented}
\vfill
\end{titlepage}
\def\thefootnote{\fnsymbol{footnote}}
\setcounter{footnote}{0}
%

\normalsize

\section{Introduction}
The Standard Model (SM) has demonstrated remarkable success in describing and predicting a wide range of phenomena, but it leaves some puzzles unsolved. One of the puzzles is the hierarchy problem, which concerns the huge gap between the electroweak (EW) scale and the scale of gravity. One class of models that attempts to address this problem, such as the composite Higgs models~\cite{composite_higgs}, predicts the existence of vector-like quarks (VLQs). The proposed heavy top quark partner T stabilizes the EW scale by canceling the quadratic divergences of the top quark loop contributions to the Higgs mass. 

The T quark is expected to decay into bW, tH, or tZ. When the mixing between the top partner and the SM top quark is small, these tree-level decays become suppressed~\cite{excited_top}. Under such conditions, loop-induced decays from another class of top quark partners, t$^* \to $ tg and t$^* \to $ t$\gamma$ dominate, where t$^*$ is the heavy excited top quark that arises in top compositeness theories~\cite{stabilize}. The masses of both classes of top partners are predicted to be on the TeV scale, within the energy reach of the LHC. This proceeding summarizes the recent CMS searches of vector-like T quark and excited top quark using full Run 2 data. 

\section{Searches for VLQs}
VLQs are non-chiral, colored, spin-1/2 fermions whose masses do not originate from Yukawa couplings, allowing them to evade constraints from the precision measurements of Higgs cross section. In minimal scenarios, the T quark couples exclusively to third-generation SM quarks and EW bosons, with branching fractions of T $\to$ bW, T $\to$ tH, and T $\to$ tZ dependent on coupling values. In non-minimal scenarios~\cite{nonminimal_vlq}, T quark decays to a top quark and a beyond SM scalar boson $\phi$ with 100\% branching fraction.

\subsection{\texorpdfstring{T $\to$ t$\phi$(H)}{T -> t phi (H)}
  in the single-lepton final state}
\begin{figure}[!htb]
\centering
\includegraphics[height=2in]{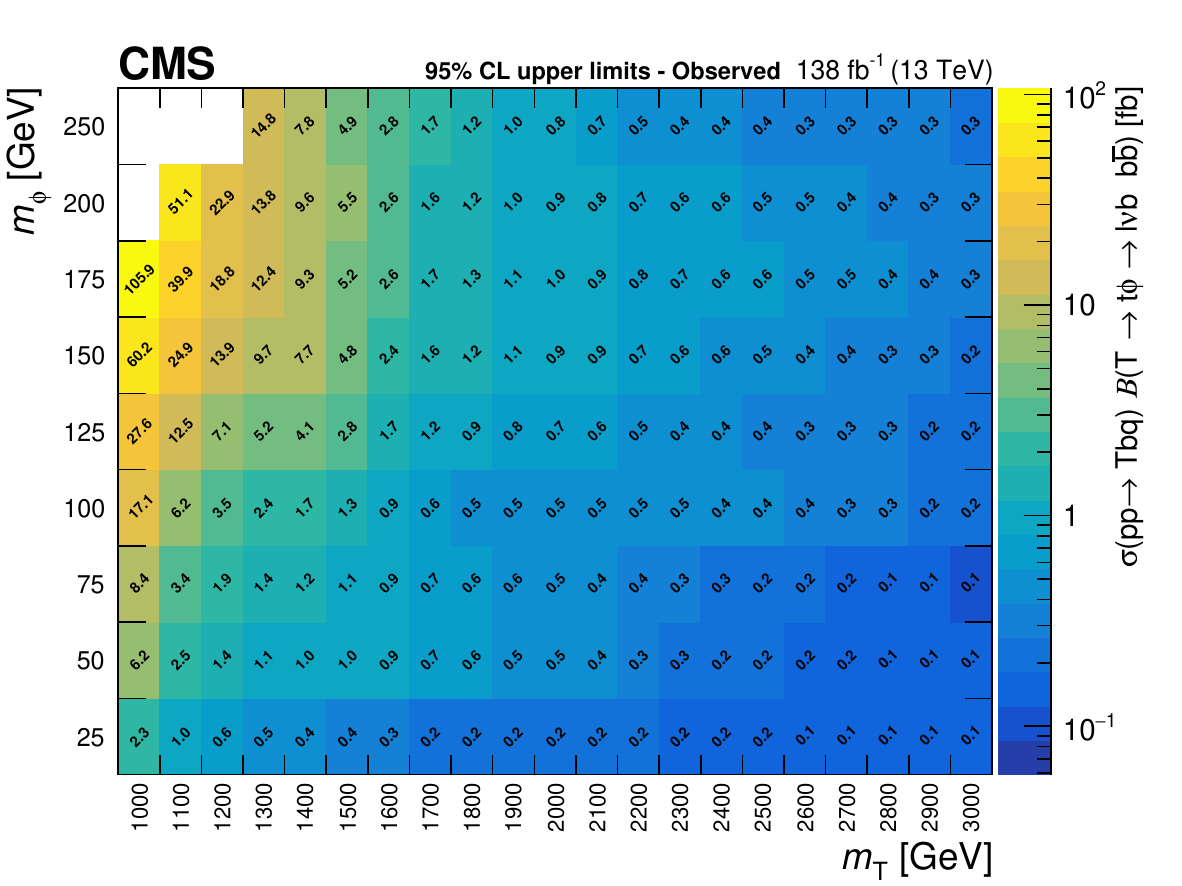}
\caption{Observed 95\% CL upper limits on the single T quark production cross section times the branching fraction of T $\to$ t$\phi$ $\to$ b$l \nu \mathrm{b \bar{b}}$ as a function of T mass and $\phi$ mass. Figure from Ref.~\cite{b2g-23-009}}
\label{fig:B2G-23-009_tphi}
\end{figure}

\begin{figure}[!htb]
\centering
\includegraphics[height=2in]{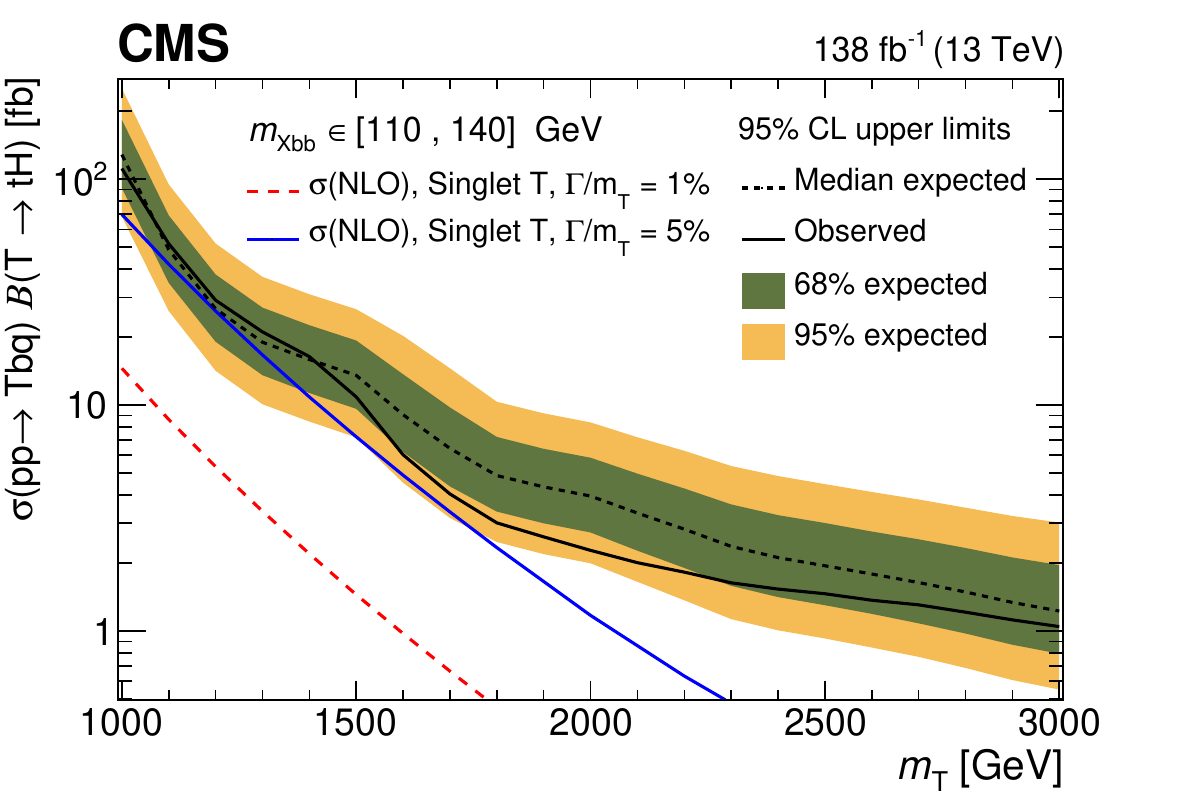}
\caption{Observed (solid black line) and expected (dashed line) 95\% CL upper limits on the single T quark production cross section times the branching fraction of T $\to$ tH. Figure from Ref.~\cite{b2g-23-009}}
\label{fig:B2G-23-009_tH}
\end{figure}

The CMS experiment has performed a search for the single production of T $\to$ t$\phi$(H) with the full Run 2 data~\cite{b2g-23-009}. The analysis searches T mass ranges of 1 -- 3 TeV and $\phi$ mass range of 25 -- 250 GeV in the single-lepton final state, with t decaying leptonically
, and $\phi$ or H decaying into a pair of b jets. This analysis exploits \textsc{ParticleNet-MD}, a mass-decorrelated algorithm based on a dynamic graph convolutional neural network~\cite{particleNet}, for H($\phi$) $\to$ b$\mathrm{\bar{b}}$ identification. This analysis also developed a boosted decision tree (BDT) top tagger that significantly improves the top identification. With these strong taggers, this analysis improves all previous CMS searches for T mass above 1.3 TeV, while maintaining comparable sensitivity down to 1.1 TeV. No significant excess is observed. Figure~\ref{fig:B2G-23-009_tphi} shows the 95\% confidence level (CL) upper limits on the product of the single T quark production cross section and the branching fraction $\mathcal{B}$(T $\to$ t$\phi$) as a function of $\phi$ mass and T mass. For the lowest (highest) searched $\phi$ mass of 25 (250) GeV, values greater than 2.0 to 0.15 fb (15 to 0.25 fb) are excluded at 95\% CL for the full searched T mass range of 1 -- 3 TeV. Figure~\ref{fig:B2G-23-009_tH} shows the corresponding limits for T $\to$ tH, excluding cross sections above 100 to 1.0 fb across the same T mass range. 

\subsection{\texorpdfstring{T $\to$ t$\phi$(H)}{T -> t phi (H)} in the hadronic final state}
\begin{figure}[!htb]
\centering
\includegraphics[height=2in]{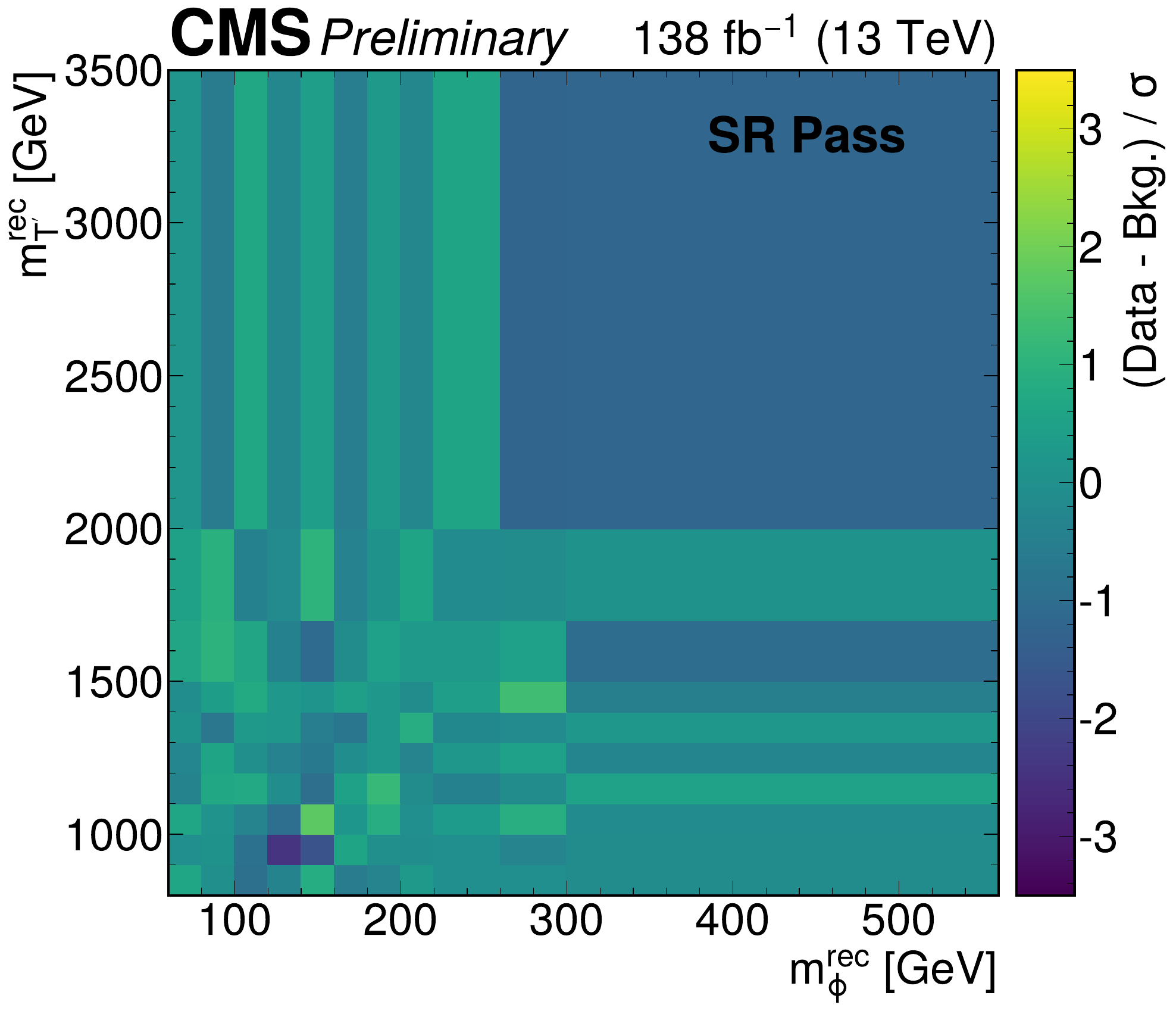}
\caption{Pull in the reconstructed $\phi$ mass and T mass in the signal region after maximum likelihood fit to the data. Figure from Ref.~\cite{CMS-PAS-B2G-22-001}}
\label{fig:B2G-22-001_pull}
\end{figure}

\begin{figure}[!htb]
\centering
\includegraphics[height=2in]{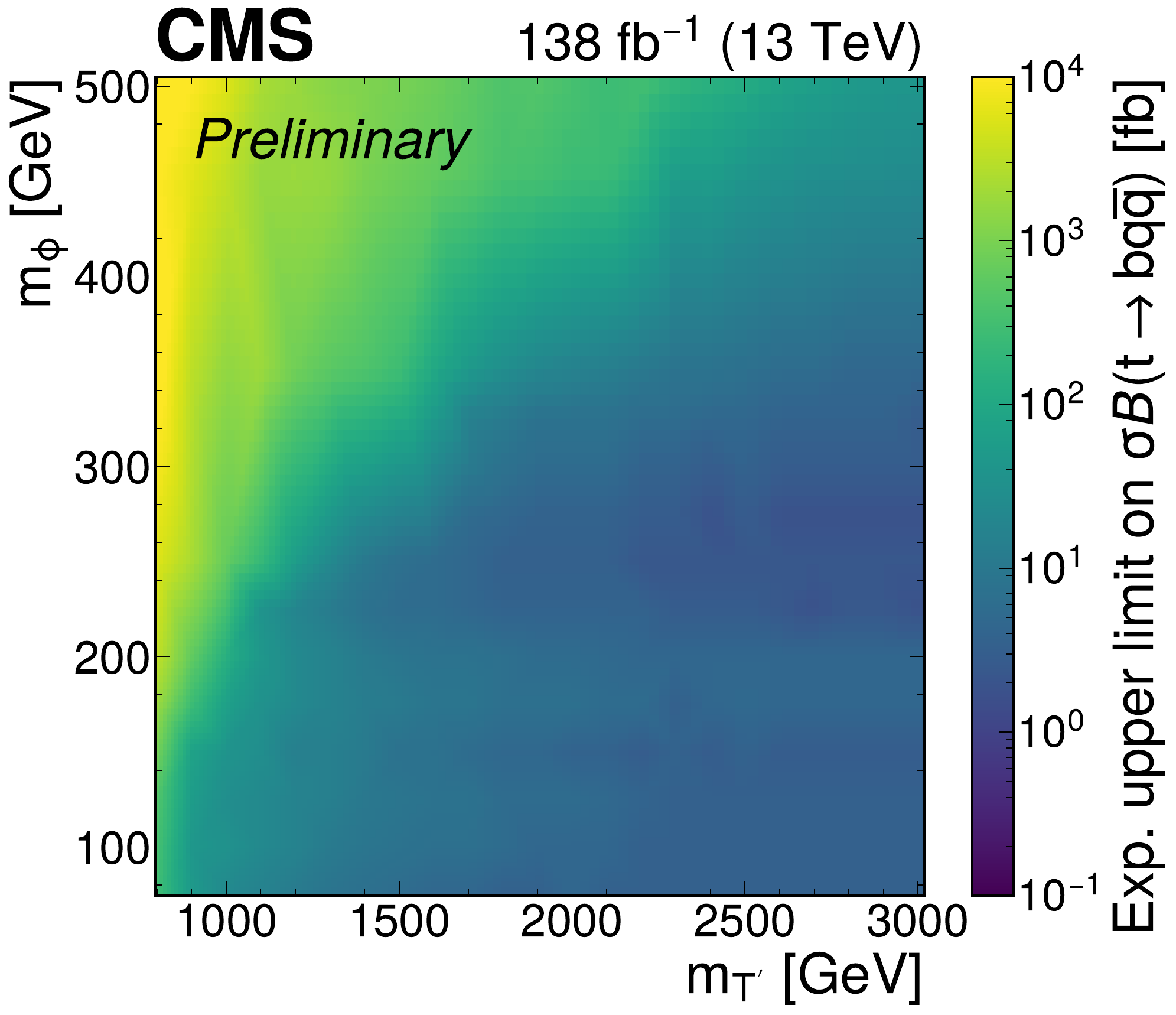}
\includegraphics[height=2in]{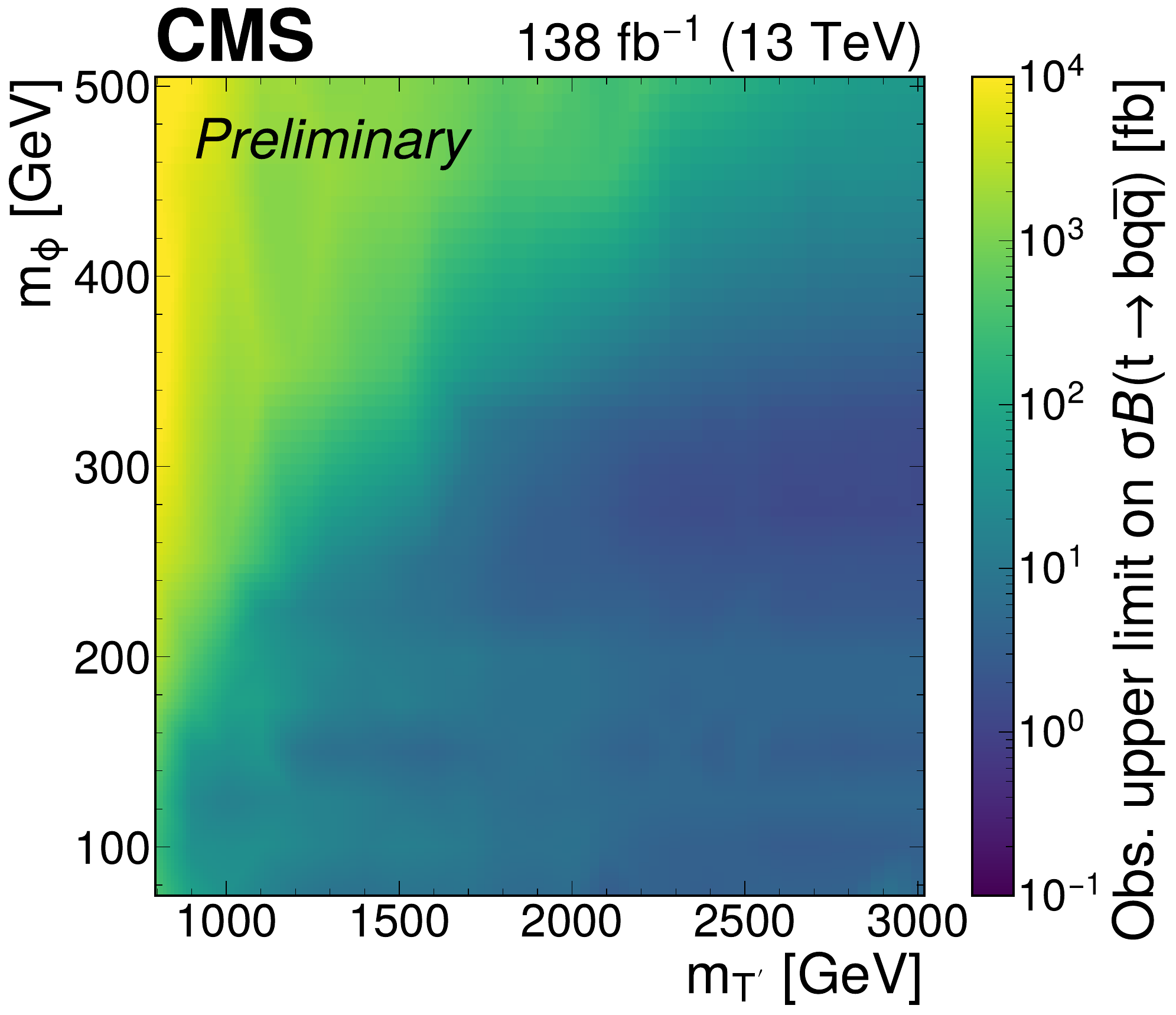}
\caption{Median expected (left) and observed (right) upper limits at the 95\% CL on the product of the T $\to$ t$\phi$ production cross section and the branching fraction $\mathcal{B}(\mathrm{t \to b q \bar{q}})$ as a function of T mass and $\phi$ mass. Figure from Ref.~\cite{CMS-PAS-B2G-22-001}}
\label{fig:B2G-22-001_2Dlimits}
\end{figure}

\begin{figure}[!htb]
\centering
\includegraphics[height=2in]{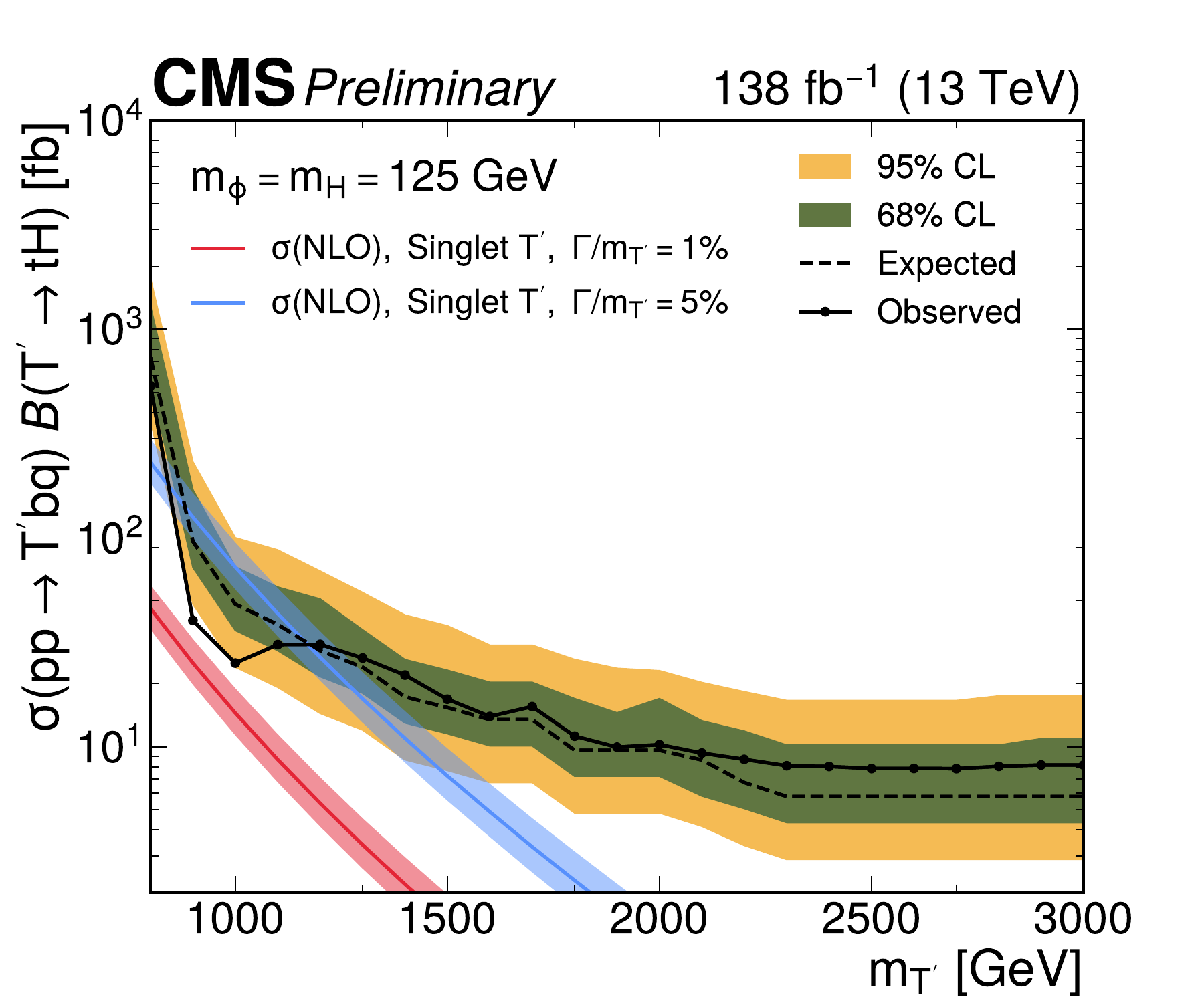}
\caption{Upper limits at 95\% CL on the product of the production cross section of T $\to$ tH and $\mathcal{B}$(T $\to$ tH) as a function of T mass for $\phi$(H) mass 125 GeV. Figure from Ref.~\cite{CMS-PAS-B2G-22-001}}
\label{fig:B2G-22-001_125Limits}
\end{figure}

In addition to single-lepton final state, CMS also performed a search of T $\to$ t$\phi$(H) in the hadronic final state~\cite{CMS-PAS-B2G-22-001}, targeting T masses in the range of 0.8 -- 3 TeV and $\phi$ masses in the range of 75 -- 500 GeV, with $\phi$(H) decaying to a pair of b quarks and t decaying into a b quark and a pair of boosted quarks. Results are extracted using a 2D binned maximum likelihood fit to the reconstructed T mass ($\mathrm{m_T}$) and $\phi$ mass ($\mathrm{m_\phi}$). Figure~\ref{fig:B2G-22-001_pull} shows the post-fit pull distribution, with a small deficit near ($\mathrm{m_T}$, $\mathrm{m_\phi}$) = (900 GeV, 125 GeV) from statistical fluctuations, which leads to a smaller observed value than the expected near these mass points in the 95\% CL upper limits on the product of production cross section and the branching fraction $\mathcal{B}(\mathrm{t \to b q \bar{q}})$ in Fig.~\ref{fig:B2G-22-001_2Dlimits}. The 95\% CL upper limits for $m_\phi =$ 125 GeV is presented in Fig.~\ref{fig:B2G-22-001_125Limits}. No significant excess is observed. The 95\% CL upper limits are set for the values of 300 to 4.6 fb for T masses between 0.8 and 3 TeV. The T quark masses below 1.2 TeV are excluded under the weak-isospin singlet scenario with a 5\% decay width relative to its mass.

\subsection{\texorpdfstring{Y/T $\to$ Wb}{Y/T -> Wb} in the single-lepton final state}
\begin{figure}[!htb]
\centering
\includegraphics[height=2in]{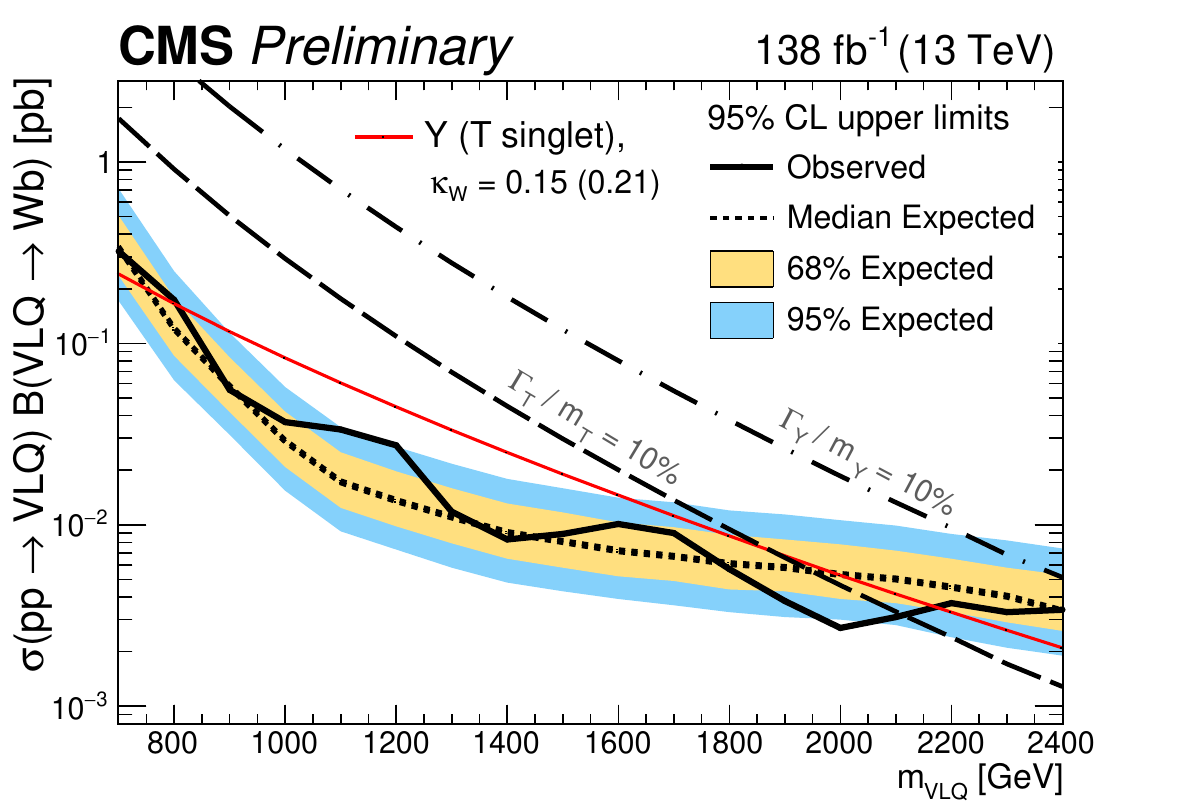}
\includegraphics[height=2in]{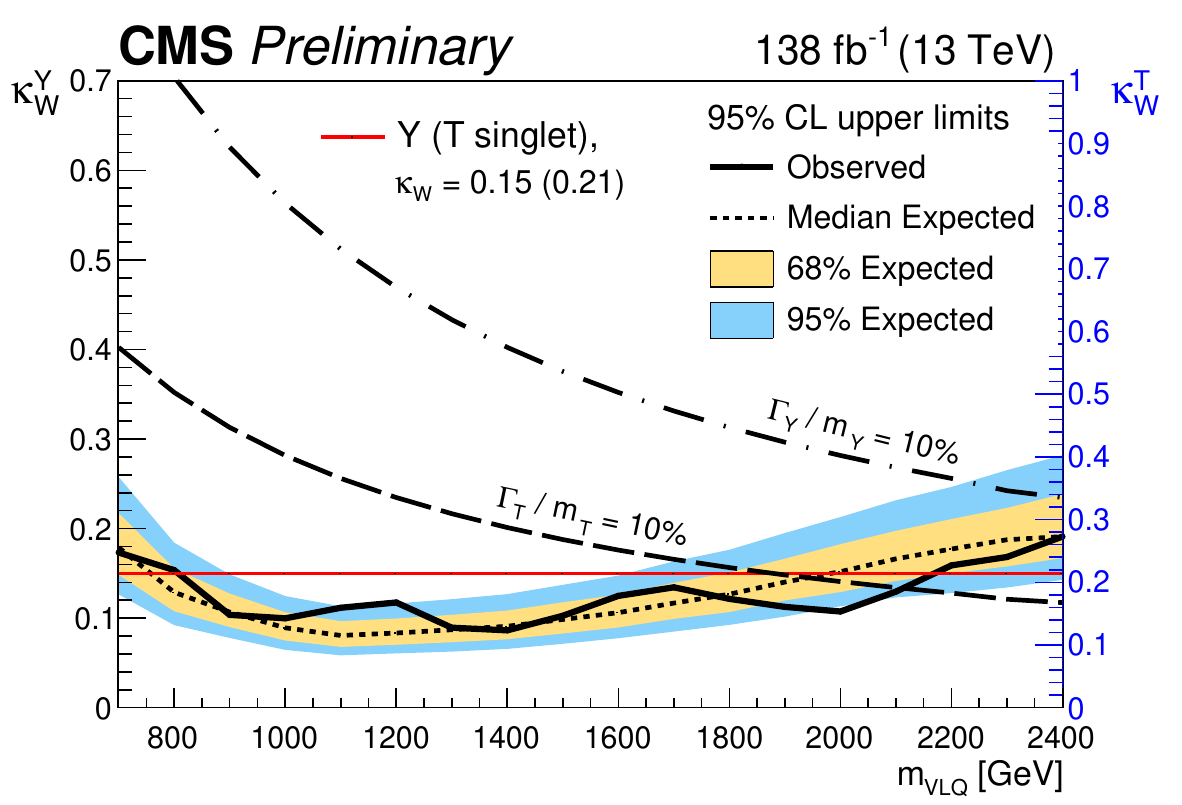}
\caption{Upper limits at 95\% CL on the product of the production cross section and $\mathcal{B}$ of Y/T $\to$ Wb (left) and coupling parameter $\kappa_W$ (right) as a function of the VLQ mass. Figure from Ref.~\cite{CMS-PAS-B2G-22-004}}
\label{fig:B2G-22-004_limits}
\end{figure}

VLQ may carry either the top-quark-like charge +2/3 (labeled as T) or an exotic charge of -4/3 (labeled as Y). Since there are no other non-SM low-mass states of Y, it decays exclusively to Wb. The CMS collaboration has conducted a search of Y/T $\to$ Wb in the single-lepton final state using the full Run 2 data~\cite{CMS-PAS-B2G-22-004}. The signal event features in a single lepton, large missing transverse momentum $\mathrm{p_T^{miss}}$, a high $\mathrm{p_T}$ b jet and forward jets. An accurate reconstruction of Y/T mass with the recursive jigsaw algorithm~\cite{jigsaw} and the W-charge asymmetry are exploited for improved sensitivity. Events in the signal region are split based on the sign of the lepton charge. Figure~\ref{fig:B2G-22-004_limits} shows the 95\% CL upper limits on the product of the production cross section and the branching fraction $\mathcal{B}$(Y/T $\to$ Wb), and the limits on the coupling parameter $\kappa_W$. Assuming $\mathcal{B}$(Y/T $\to$ Wb)=100\% and $\kappa_W$ between 0.2 and 0.15, Y(T) masses are excluded in the range of 0.7 -- 2.4 (0.82 -- 2.15) TeV.

\section{Searches for excited fermions}
Excited top quarks t$^*$ are predicted to carry the same weak isospin, color and weak hypercharge as the SM top quark. They may exist as spin-1/2 or spin-3/2 fermions. The dynamics of spin-1/2 t$^*$ is described in a similar way as other heavy SM quarks~\cite{excited_top}. The dynamics of spin-3/2 t$^*$ is described by a Rarita-Schwinger Lagrangian, which adapts the Dirac equation to a spin-3/2 particle~\cite{excitedt_spin32}. The t$^*$ quarks can decay to tg and t$\gamma$, with $\mathcal{B}$(t$^* \to$ tg) $=$ 97\% and $\mathcal{B}$(t$^*$ $\to$ t$\gamma$) $=$ 3\%.

\subsection{\texorpdfstring{$\mathrm{t^* \bar{t}^{*} \to tg\bar{t}g}$}{t* tbar* -> t g tbar g}}
\begin{figure}[!htb]
\centering
\includegraphics[height=2in]{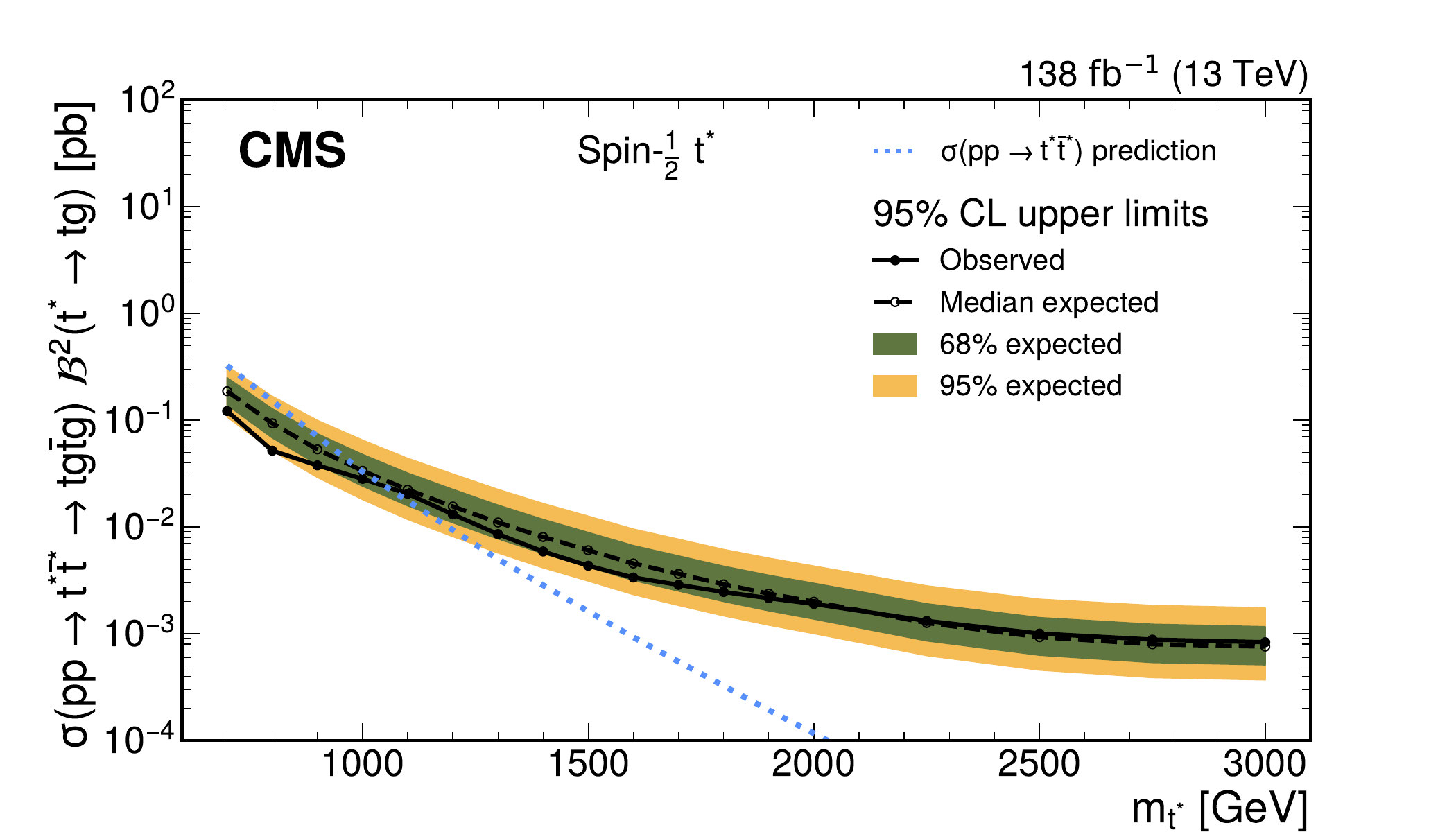}
\includegraphics[height=2in]{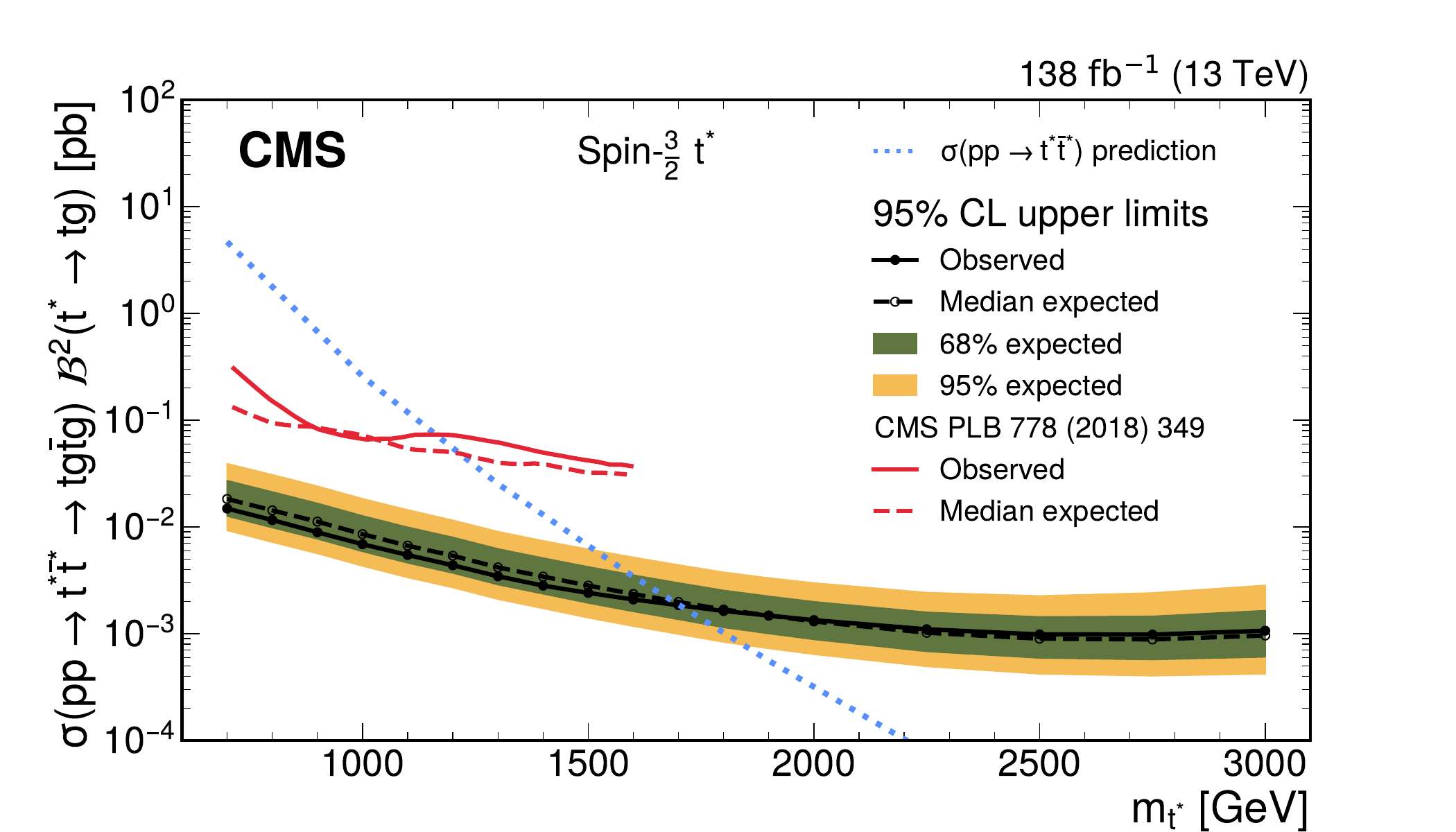}
\caption{Upper limits at 95\% CL on the product of the production $\mathrm{t^* \bar{t}^*}$ cross section and the branching fraction squared $\mathcal{B}^2(\mathrm{t^* \to tg})$ for spin-1/2 (upper) t$^*$ and spin-3/2 (lower) t$^*$. Figure from Ref.~\cite{b2g-22-005}}
\label{fig:B2G-22-005_limits}
\end{figure}

The high branching fraction of $\mathrm{t^* \to tg}$ motivates a search of the pair production of t$^*$ in the tgtg final states. The CMS collaboration has performed a search of $\mathrm{t^* \bar{t}^* \to tg\bar{t}g}$ using full Run 2 data, with one of the top quark decaying leptonically and the other decaying hadronically~\cite{b2g-22-005}. The final states feature in multiple light jets and two b jets. This analysis uses variable-radius jets to reconstruct the t$^*$ candidates and performs a binned maximum likelihood fit on $\mathrm{S_T}$, which is the $\mathrm{p_T}$ sum of the signal lepton, $\mathrm{p_T^{miss}}$, and the variable-size jets. A deep neural network (DNN) signal-to-background discriminator is used with its score decorrelated with $\mathrm{S_T}$ by the designing decorrelated taggers (DDT) technique~\cite{ddt}. This analysis substantially improves the previous CMS result from analyzing data with an integrated luminosity of 35.9 fb$^{-1}$, through a bigger dataset and improved analysis techniques. Figure~\ref{fig:B2G-22-005_limits} shows the expected and observed 95\% CL upper limits on the production $\mathrm{t^* \bar{t}^*}$ cross section and the branching fraction squared $\mathcal{B}^2(\mathrm{t^* \to tg})$ for spin-1/2 and spin-3/2 t$^*$. Spin-1/2 (Spin-3/2) t$^*$ is excluded up to mass of 1050 (1700) GeV. This represents the first exclusion limit on a spin-1/2 t$^*$ in the t$^* \to$ tg decay channel. 

\subsection{\texorpdfstring{$\mathrm{t^* \bar{t}^* \to tg\bar{t} \gamma}$}{t* tbar* -> t g tbar gamma}}
\begin{figure}[!htb]
\centering
\includegraphics[height=2in]{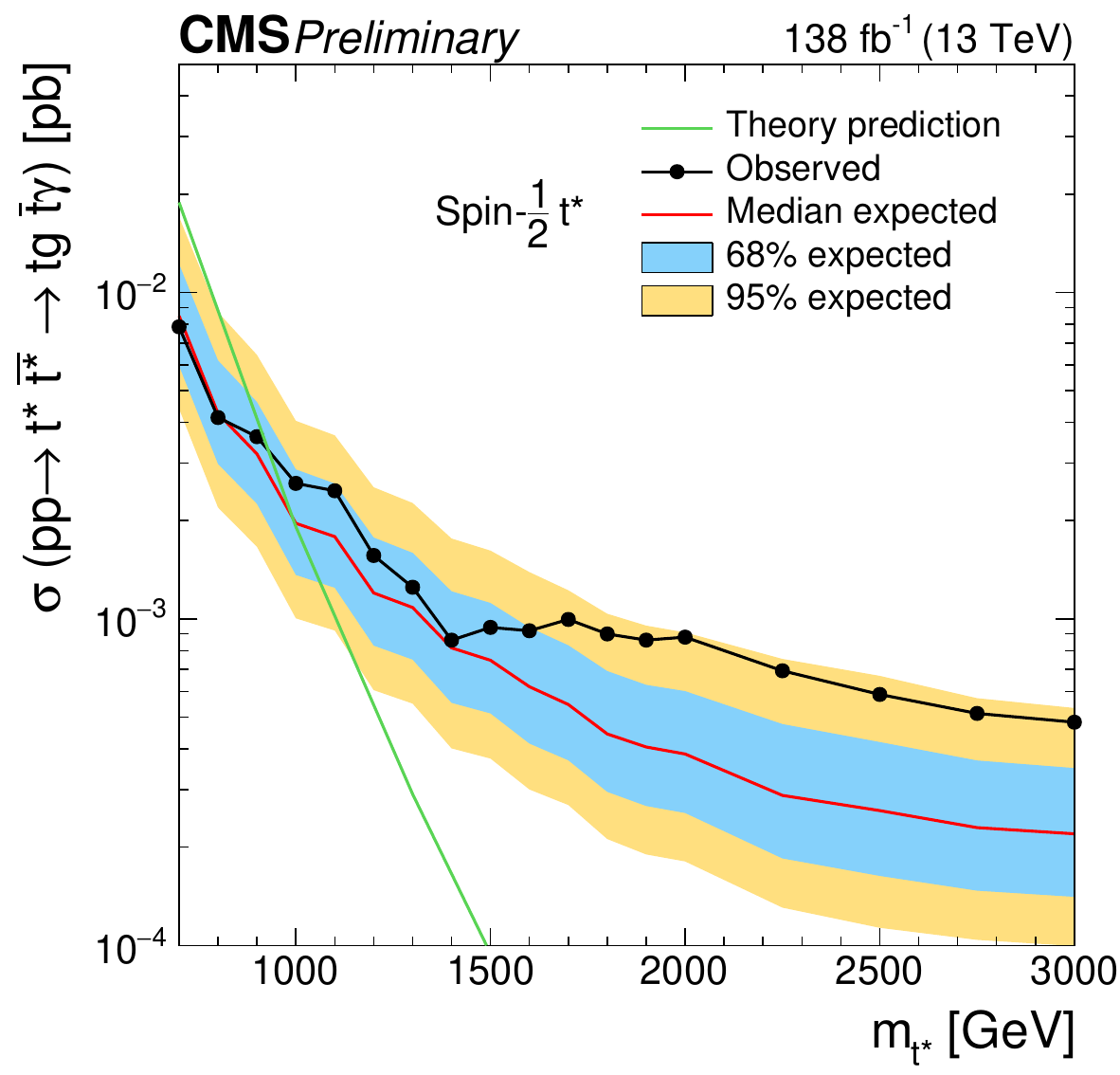}
\includegraphics[height=2in]{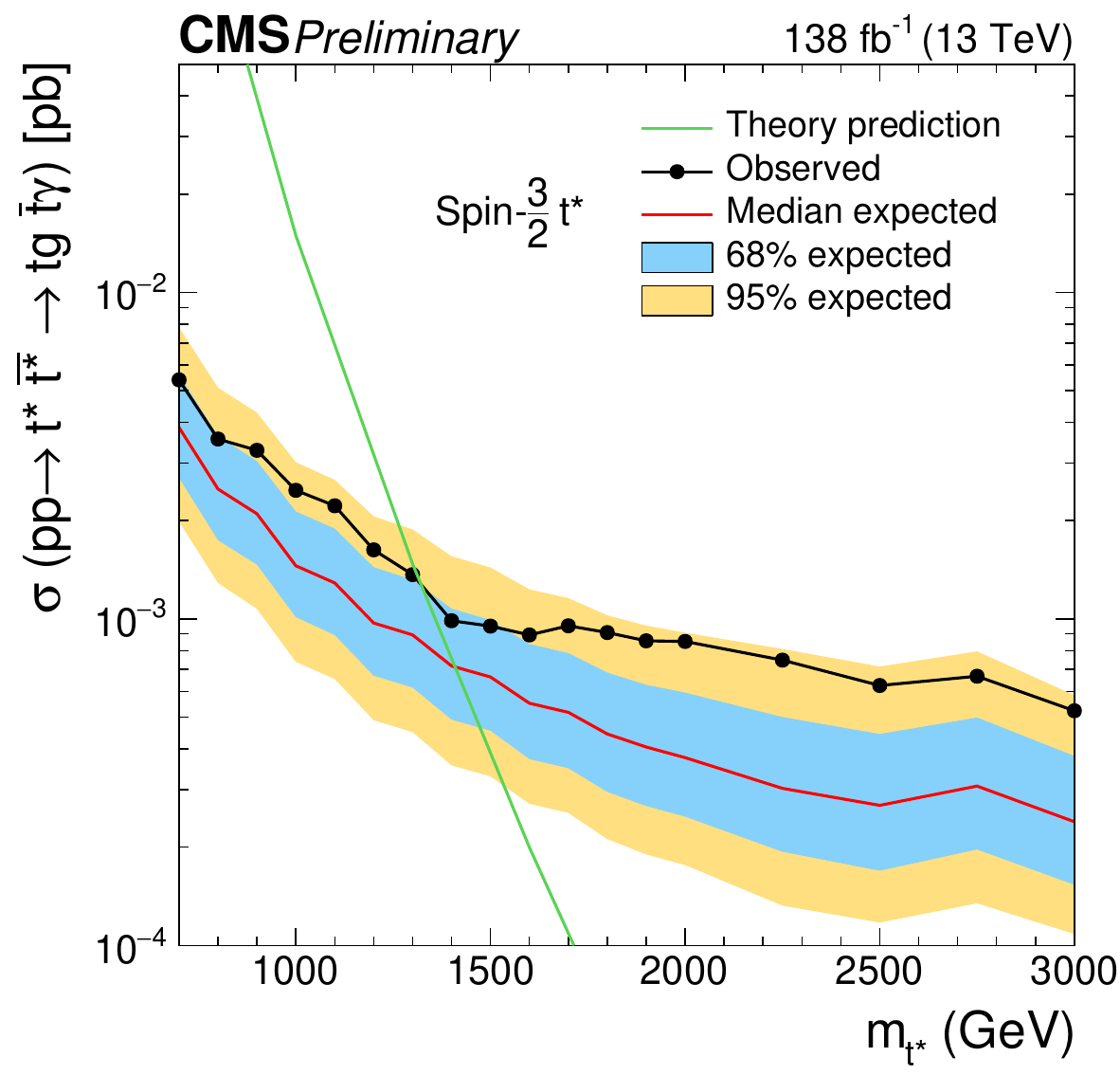}
\caption{Upper limits at 95\% CL on the product of the production $\mathrm{t^* \bar{t}^*}$ cross section and the branching fractions $\mathcal{B}(\mathrm{t^* \to tg}) \mathcal{B}(\mathrm{t^* \to t \gamma})$ for spin-1/2 (left) t$^*$ and spin-3/2 (right) t$^*$. Figure from Ref.~\cite{CMS-PAS-B2G-24-006}}
\label{fig:B2G-24-006_limits}
\end{figure}

Although the branching fraction of $\mathrm{t^* \to t \gamma}$ is onl 3\%, the clean signature provided by the high-momentum photon makes $\mathrm{t^* \bar{t}^* \to tg\bar{t} \gamma}$ another highly motivated channel. The CMS collaboration has searched for $\mathrm{t^* \bar{t}^* \to tg\bar{t} \gamma}$ with the full Run 2 data~\cite{CMS-PAS-B2G-24-006}. The signal event features in a high transverse momentum photon and multiple large radius jets from hadronic top decays. The $t^*$ mass is reconstructed from t$\gamma$. SM background is significantly suppressed by exploiting the properties of high-energy photon. Remaining background is categorized based on photon origin, into ``prompt photon", ``jet misidentified as photon", and ``Electron misidentified as photon". This analysis achieves comparable sensitivity with the $\mathrm{t^* \bar{t}^* \to tg\bar{t}g}$ search. It excludes spin-1/2 (spin-3/2) t$^*$ for masses up to 940 (1330) GeV. The observed and expected 95\% CL upper limits are shown in Figure~\ref{fig:B2G-24-006_limits}.

\section{Conclusions}
Recent results on VLQ and excited fermions searches are presented in this proceeding. No significant excess has been observed. The full Run 2 results significantly improved proceeding results through a bigger dataset and through novel techniques and analysis strategies, such as advanced jet taggers and dedicated reconstruction techniques.

\clearpage
\bibliographystyle{plain}
\bibliography{bibliography}

\end{document}